\documentclass[pra,11pt,superscriptaddress]{revtex4-1}

\usepackage{amsmath}   
\usepackage{bm}        
\usepackage{graphicx}  
\usepackage{verbatim}  
\usepackage{color}     
\usepackage{hyperref}  
\usepackage{bm}
\usepackage{multirow}
\usepackage{longtable}
\newcommand{\lb}{ \langle}
\newcommand{\rb}{ \rangle}

\begin{document}

\title{The effect of an external magnetic field on the determination of E1M1 two-photon decay rates in Be-like ions}

\author{Jon Grumer}\email{jon.grumer@teorfys.lu.se}
\affiliation{Division of Mathematical Physics, Department of Physics, Lund University, S-221 00 Lund, Sweden}

\author{Wenxian Li}
\affiliation{Institute of Modern Physics, Fudan University, 200433 Shanghai, China}

\author{Dietrich Bernhardt}
\affiliation{Institut f\"{u}r Atom- und Molek\"{u}lphysik, Justus-Liebig-Universit\"{a}t Giessen, 35392 Giessen, Germany}

\author{Jiguang Li}
\affiliation{Division of Mathematical Physics, Department of Physics, Lund University, S-221 00 Lund, Sweden}

\author{Stefan Schippers}
\affiliation{Institut f\"{u}r Atom- und Molek\"{u}lphysik, Justus-Liebig-Universit\"{a}t Giessen, 35392 Giessen, Germany}

\author{Tomas Brage}	
\affiliation{Division of Mathematical Physics, Department of Physics, Lund University, S-221 00 Lund, Sweden}

\author{Per J\"{o}nsson}
\affiliation{Group for Materials Science and Applied Mathematics, Malm\"{o} University, S-205 06 Malm\"{o}, Sweden}

\author{Roger Hutton}
\affiliation{Institute of Modern Physics, Fudan University, 200433 Shanghai, China}

\author{Yaming Zou}
\affiliation{Institute of Modern Physics, Fudan University, 200433 Shanghai, China}

\date{\today}

\begin{abstract}
In this work we report on \textit{ab initio} theoretical results for the magnetic field induced $2s\,2p~^3P_0~\to~2s^2~^1S_0$ E1 transition for ions in the beryllium isoelectronic sequence between $Z=5$ and $92$. 

It has been proposed that the rate of the E1M1 two-photon transition $2s\,2p~^3P_0~\to~2s^2~^1S_0$ can be extracted from the lifetime of the $^3P_0$ state in Be-like ions with zero nuclear spin by employing resonant recombination in a storage-ring. This experimental approach involves a perturbing external magnetic field. The effect of this field needs to be evaluated in order to properly extract the two-photon rate from the measured decay curves.

The magnetic field induced transition rates are carefully evaluated and it is shown that, with a typical storage-ring field strength, it is dominant or of the same order as the E1M1 rate for low- and mid-$Z$ ions. Results for several field strengths and ions are presented and we also give a simple $Z$-dependent formula for the rate. We estimate the uncertainties of our model to be within $5\%$ for low- and mid-Z ions, and slightly larger for more highly charged ions. Furthermore we evaluate the importance of including both perturber states, $^3P_1$ and $^1P_1$, and it is shown that excluding the influence of the $^1P_1$ perturber overestimates the rate by up to $26\%$ for the mid-$Z$ ions.
\end{abstract}

\maketitle

\section{Introduction \label{sec:intro}}

Two-photon transitions are exotic decay modes in atoms and ions. Nevertheless, they are of practical interest, e.g. in astrophysics  where the $2s\to1s$ (2E1) transition in hydrogen contributes to the observed continuum radiation from planetary nebulae \cite{Spitzer1951}, Herbig-Haro objects \cite{Dopita1982}, and H\,II regions \cite{Osterbrock2006}. Theoretical work on two-photon transitions started already at the dawn of quantum mechanics  \cite{GoeppertMayer1931}. Since then, theoretical and experimental work has mainly focussed on H-like and He-like systems \cite[and references therein]{Mokler2004}. Various aspects of two-photon transitions, such as resonance effects \cite{Amaro2009}, negative continuum effects \cite{Surzhykov2009}, relativistic and QED effects \cite{Wundt2009}, and higher-order multipole effects \cite{Solovyev2010} on two-photon transitions in H-like ions, in these isoelectronic sequences of ions have been addressed in very recent (mostly theoretical) studies. Additionally, the sensitivity of the spectral shape of the emitted photon continuum to relativistic effects \cite{Trotsenko2010}, as well as angular correlations  \cite{Surzhykov2010} and quantum correlations \cite{Fratini2010} between the two-photons have been investigated.

In He-like ions  there exist three long-lived metastable states which decay (partly) via the two-photon transitions  $1s\,2s~^1S_0~\to~1s^2~^1S_0$ (2E1), $1s\,2s~^3S_1~\to~1s^2~^1S_0$ (2E1), and  $1s\,2p~^3P_0~\to~1s^2~^1S_0$ (E1M1). The latter competes with the dominating $1s\,2p~^3P_0~\to~1s\,2s~^3S_1$ one-photon E1 transition. The relative importance of the E1M1 transition increases with nuclear charge $Z$. For $Z=92$ the E1M1 branching ratio has been calculated to amount to about 32\% \cite{Savukov2002,Labzowsky2004}.

With Be-like ions the situation is much more clear-cut. The $2s\,2p~^3P_0$ state is the lowest excited state and for isotopes with a non-zero nuclear spin, the $J = 0 \to 0$ transition channel opens up due to mixing of the hyperfine levels, leading to a so-called hyperfine induced transition (HIT). Such a shortening of lifetimes of metastable states owing to hyperfine interaction, is referred to as hyperfine quenching and has been investigated for Be-like ions both theoretically \cite{Brage1998, Marques1993, ChengPRA2008, AnderssonPRA2009, Li2010, Li2011} and experimentally \cite{SchippersPRL2007, SchippersPRA2012}. For isotopes with zero nuclear spin on the other hand, a one-photon transition to the ground state $2s^2~^1S_0$ is strictly forbidden in a field-free region and the lowest-order decay channel is a very slow E1M1 two-photon process. The most important third-order process is a 3E1 three-photon decay, which has a transition rate smaller than the two-photon process by a factor of $\alpha$, the fine structure constant, according to Laughlin \cite{LaughlinPLA1980}.
   
The calculation of E1M1 rates involves  potentially significant negative-energy contributions to the transition amplitudes \cite{Savukov2002}. Thus, an accurate measurement of the experimental decay rate would constitute an ultimate benchmark of relativistic many-body theoretical methods and computational schemes. So far there exist no experimental observations of E1M1 transitions in He-like or Be-like systems \cite{Mokler2004}.

Recently, future storage-ring experiments have been proposed \cite{Schippers2011c,Bernhardt2012,SchippersArXiv2012} to measure $2s\,2p~^3P_0~\to~2s^2~^1S_0$ E1M1 two-photon transition rates for heavy Be-like ions with nuclear charges $Z \gtrsim 50$. However, the magnetic field of the storage-ring dipole magnets will give rise to a magnetic field induced E1 transition (from here on referred to as a MIT), possibly with a rate of the same order of magnitude as the rate of the two-photon transition. Hence, to correctly deduce the two-photon rate from such an experiment, there is a need for accurate MIT rates as discussed in \cite{SchippersArXiv2012}.

Arguably a MIT was observed for the first time in 2003 by Beiersdorfer \textit{et al.} in Ne-like Ar, using the EBIT-II electron beam ion trap at the Lawrence Livermore National Laboratory \cite{BeiersdorferPRL2003}. They also showed that such transitions can play an important role in high-temperature plasma diagnostics, e.g. in fusion reactors.

In the present work we investigate the mixing, as induced by an external magnetic field, of the four atomic fine-structure states $2s\,2p~^{1,3}P_{0,1,2}$ in Be-like ions with zero nuclear spin, giving rise to a MIT from $^3P_0$ to the ground state. Note that there will also be a MIT from $^3P_2$ to the ground state, which is of no direct interest to this work.

In the following section (\ref{sec:rings}) we introduce the relevant details of storage-ring measurements of atomic lifetimes. Our theoretical methods are described in section \ref{sec:method} and computational details are given in section \ref{sec:model}. Our results and conclusions are presented in sections \ref{sec:results} and \ref{sec:conc}.

\section{Storage-ring measurements of atomic lifetimes}\label{sec:rings}

Heavy-ion storage-rings are ideal devices for measuring atomic lifetimes \cite{Traebert2010}. They provide a unique experimental environment, which is characterized by low residual gas density and correspondingly long ion storage times of up to several hours \cite{Grieser2012}. In a typical experiment, ions with well defined charge state, mass, and kinetic energy are injected into the storage-ring from an external accelerator. Beam cooling techniques such as electron cooling \cite{Poth1990} or stochastic cooling \cite{Caspers2012} may be applied to reduce internal energy spread and diameter of the stored ion beam, i.e, to create well controlled experimental conditions. Long-lived metastable levels of interest are usually generated by the charge stripping process that is used for producing the desired ion charge state in the accelerator. Metastable levels may also be populated in situ by collisional excitation \cite{Mannervik1997a} or by optical pumping \cite{Mannervik1999a}.

The standard technique for measuring atomic decay rates in an ion storage-ring is to monitor the fluorescence from the long-lived excited levels as function of storage time \cite{Traebert2010}. However, this approach suffers from small solid angles and background photons which severely hampers the investigation of weak decay channels \cite{Traebert2011}. An alternative approach is electron-ion collision spectroscopy, where level-specific charge changing electron-ion collision processes such as dielectronic recombination (DR) are exploited for monitoring the decay of the metastable ion beam fraction \cite{Schmidt1994}. This techniques yields comparatively high signal rates since the fast moving product ions are confined into a narrow cone and can thus be detected with high efficiency. It has been successfully employed for the measurement of $2s\,2p~^3P_0~\to~2s^2~^1S_0$ HIT rates in Be-like $^{47}$Ti$^{18+}$ \cite{SchippersPRL2007} and $^{33}$S$^{12+}$ \cite{SchippersPRA2012} at the Heidelberg heavy-ion storage-ring TSR.
It has been proposed \cite{Schippers2011c,Bernhardt2012,SchippersArXiv2012} to use the same technique  at the heavy-ion storage-ring ESR of the GSI Helmholtz Center for Heavy-Ion Research in Darmstadt, Germany for the measurement of the $2s\,2p~^3P_0~\to~2s^2~^1S_0$ E1M1 two-photon transition rates in heavy Be-like ions with zero nuclear spin.

An issue of concern in these measurements is in how far the atomic lifetimes are influenced by the magnetic fields that are generated by the storage-ring dipole, quadrupole, and higher order correction magnets.
Magnetic quenching in the TSR has been investigated theoretically by Li et al. \cite{Li2011} for the $2s\,2p~^3P_0$ level in Be-like $^{47}$Ti$^{18+}$. The effect has been found to be insignificant for this specific case.
The effect of external fields on the HIT rates has also been investigated experimentally. In the S$^{12+}$ storage-ring experiment \cite{SchippersPRA2012} the magnetic field strength of the storage-ring dipoles was varied by a factor of two. Within the experimental uncertainties no influence of this $B$-field variation on the measured HIT rate was found. In contrast to the HIT and E1M1 transition rates, which in general are increasing with $Z$, the MIT rate decreases with $Z$. Hence it can be expected that the relative importance of magnetic quenching decreases with increasing $Z$.

\begin{figure}
\center
\includegraphics[width=0.4\textwidth]{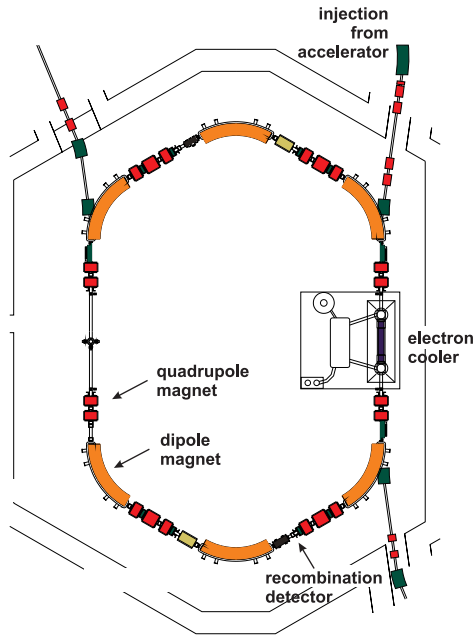}
\caption{\label{fig:1a} Sketch of the heavy-ion storage-ring ESR of the GSI Helmholtz Center for Heavy-Ion Research in Darmstadt, Germany. The ESR storage-ring \cite{Franzke1987} has a circumference of 108.36~m. It has a hexagonal layout consisting of 6 dipole bending magnets, 2 long straight sections, and 4 short straight sections. The lengths of long and short straight sections are $18.235$~m and $8.155$~m, respectively. The dipole magnets have a bending angle of 60$^\circ$ and a bending radius $\rho = 6.25$~m. The length of the flight path through all dipole magnets is $2\pi\rho = 39.27$~m, i.e., the six magnets cover $36.24\%$ of the ring circumference. The maximum magnetic field $B$ is about 1.5~T.}
\end{figure}

Although the ions which move with typically 10-30\% the speed of light, $c$, are subjected to an alternating magnetic field with frequencies in the MHz range, we here  as a first approach treat the magnetic field as constant. This implies that we also neglect the magnetic fields of the quadrupole and higher order correction magnets, which are anyway much weaker than the field in the bending dipole magnets.

It should also be noted that the magnetic field transforms into an electric field $\mathbf{E} = q(\mathbf{v}\times\mathbf{B})$ in the rest frame of the ions with charge $q$ moving with velocity $\bm v$. Under rather extreme experimental conditions, i.e., for $v=c$ and $B=1.5$~T the electric field strength amounts to $4.5\times10^8$~V~m$^{-1}$. In search for parity-violating effects Maul et al.~\cite{Maul1998a} have calculated the rate for the associated quenching of the $2s\,2p\;^3P_0$ level. This rate scales quadratically with field strength, $E$. Even for $E=4.5\times10^8$~V~m$^{-1}$ the effect is very weak. The associated transition rates are smaller than $\sim2\times10^{-5}$~s$^{-1}$ \cite{SchippersArXiv2012} and, thus,  insignificant for the present study.

\section{Theoretical Method} \label{sec:method}

The details of the theoretical approach used in this work have been outlined in a recent paper on MITs in Ne-like ions \cite{LiPRA2013}. Here we just briefly summarize the method.

The Hamiltonian of an atom with zero nuclear spin under the influence of an external homogeneous magnetic field, $\mathbf{B}$, can be written in the following form \cite{ChengPRA1985}
\begin{equation}
\label{eq:Hm}
H =H_{fs} +H_m=H_{fs} + \left(\mathbf{N}^{(1)}+\Delta \mathbf{N}^{(1)}\right)\cdot\mathbf{B}
\end{equation}
where the first term, $H_{fs}$, is the relativistic fine structure Hamiltonian which in our approach includes the Breit-interaction and leading QED effects such as self-energy and vacuum polarization. The tensor operator $ \mathbf{N}^{(1)}$ represents the coupling of the electrons with the field and $\Delta \mathbf{N}^{(1)} $ is the Schwinger QED correction. Explicit forms of the operators can be found in \citep{ChengPRA1985}.

In the presence of an external magnetic field, $M$ (and parity, which we leave out for simplicity) is the only good electronic quantum number, and we expand the $M$-dependent atomic state functions $|M\rb$ in terms of field-independent atomic state functions (ASFs), $|\Gamma J M \rb$, that are eigenstates of the fine structure Hamiltonian
\begin{equation} \label{eq:4}
|M \rb = \sum_{\Gamma J} d_{\Gamma J} |\Gamma J M \rb ~.
\end{equation}
The mixing coefficients associated with the magnetic field perturbation, $d_{\Gamma J}$, can be obtained through first order perturbation theory,
\begin{equation} \label{eq:mixingcoff}
d_{\Gamma J} =  \frac{\lb \Gamma J M |H_m| \Gamma_0 J_0 M_0 \rb}{E(\Gamma_0 J_0) - E(\Gamma J)}
\end{equation}
where the labels having a subscript zero denote the reference state. Alternatively one can evaluate the mixing by solving the corresponding eigenvalue problem.

In order to construct the ASFs we use the multiconfiguration Dirac-Hartree-Fock (MCDHF) approach \cite{Grant2007}. The starting point of this method is to write the ASFs as linear combinations of configuration state functions (CSFs), which in turn are Eigenfunctions of $J^2$, $J_z$ and parity
\begin{equation} \label{eq:csf_exp}
|\Gamma JM\rb = \sum_i c_i |\gamma_i JM\rb \, ,
\end{equation}
where $c_i$ are mixing coefficients of the CSFs, and $\gamma_i$ are labels, such as orbital occupation numbers and intermediate spin-angular couplings, to uniquely define the individual basis functions. Each of these many-electron CSFs are in turn constructed as coupled antisymmetric sum of products of one-electron wavefunctions, the Dirac-orbitals.

Applying the basis expansion (\ref{eq:4}), the electric dipole transition probability for a magnetic field induced transition from an initial state $|M'\rb$ to a final state $|M\rb$ is given by
\begin{equation} \label{eq:A1}
A_{MIT} = \frac{2.02613\times10^{18}}{\lambda^3} \sum_q \Bigg|\sum_{\Gamma J} \sum_{\Gamma' J'} (-1)^{J-M}d_{\Gamma J}d'_{\Gamma' J'}
\left (
\begin{array}{ccc}
J  & 1 & J'\\
-M & q & M'
\end{array}
\right )
\left\lb \Gamma J \big|\big| \mathbf{P}^{(1)} \big|\big| \Gamma' J'\right\rb \Bigg|^2
\end{equation}
where $A_{MIT}$ is in $s^{-1}$ and $\lambda$ is the wavelength of the transition in \AA. One should keep in mind that the real photon energy, that is the transition energy of the induced transition under consideration (i.e between $2s^2~^1S_0$ and $2s\,2p~^3P_0$ in this case) must be used to calculate the electric dipole transition matrix elements~\cite{ChengPRA2008}.

The magnetic interaction induces mixing between states that differ in $J$ by at most 1, hence the regular E1 selection rule of change in total angular momentum is extended to $\Delta J = J - J' = 0, \pm 1, \pm 2, \pm 3 $. The mixing also implies that what appears as a $J=0\to 0$ transition, is allowed.

\begin{figure}[b]
\center
\includegraphics[width=0.4\textwidth]{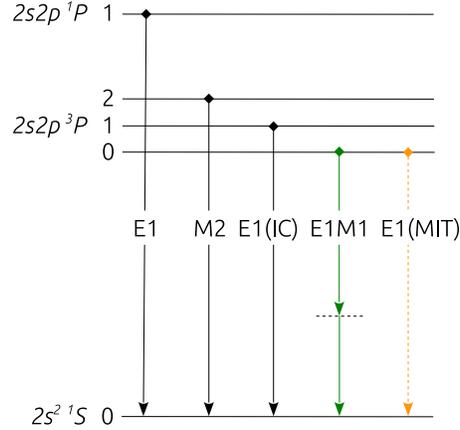}
\caption{\label{fig:1}Schematic Grotrian diagram at low $Z$, where $LS$-coupling notation is appropriate, for the lowest states of Be-like ions with zero nuclear spin. The lowest order decay from $^3P_0$ is the E1M1 two-photon transition to the groundstate.  In the presence of an external magnetic field, the usually strictly forbidden one-photon transition channel $2s\,2p~^3P_0 \to 2s^2~^1S_0$ is opened due to mixing with $^3P_1,~M=0$, which is decaying to the groundstate through the unexpected E1 intercombination channel (IC), and with $^1P_1,~M=0$, which decays to the groundstate with an allowed E1 transition.}
\end{figure}

The general theory can be applied to the MIT rates in Be-like ions. The reference state $2s\,2p~^3P_{0}$ in these systems, under the influence of an external magnetic field (see Fig. \ref{fig:1} for schematics of energy structure and possible transition channels), can approximately be expressed as
\begin{equation} \label{eq:be_like_expansion}
\left|"2s\,2p~^3P_0"~M=0 \right\rb = d_0 \left| 2s\,2p~^3P_0~M=0 \right\rb + \sum_{\mathcal{S}(=1,3)} d_{\mathcal{S};J=1} \left| 2s\,2p~^\mathcal{S}P_1~M=0 \right\rb \, ,
\end{equation}
where further interactions have been excluded due to large energy separations and relatively weak magnetic interaction couplings. The quotation marks are used to clarify that the notation is just a label corresponding to the largest $J$-dependent eigenvector component. The ground state is more or less isolated from other states, so the corresponding $M$-dependent state is very well described by a single ASF
\begin{equation}
\left|" 2s^2~^1S_{0}"~M=0 \right\rb = \left| 2s^2~^1S_{0}~M=0 \right\rb ~ .
\end{equation}

The inclusion of the perturbing states $\left| 2s\,2p~^{1,3}P_{1}~M=0 \right\rb$ in the wavefunction, Eq. (\ref{eq:be_like_expansion}), opens up one-photon E1 transitions to the ground state. Using Eq. (\ref{eq:A1}) and evaluating the 3-j symbol, the corresponding transition rates can be expressed as
\begin{equation} \label{MIT}
A_{MIT} = \frac{2.02613\times10^{18}}{3\lambda^3}\Bigg| \sum_{\mathcal{S}(=1,3)} d_\mathcal{S} \left\lb 2s\,2p~^1S_0 || \mathbf{P}^{(1)} || 2s\,2p~^{S}P_1 \right\rb \Bigg|^2 ~ .
\end{equation}

Finally, since the mixing coefficients $d_{\mathcal{S}}$ in first order perturbation theory, are directly proportional to the magnetic field strength, we define a reduced mixing coefficient $d_\mathcal{S}^R$ and hence also a reduced transition rate, $A_{MIT}^R$, which in effect are independent of $B$, through
\begin{equation} \label{eq:RMIT}
d_\mathcal{S} =  B\, d_\mathcal{S}^R ~, \qquad A_{MIT} =  B^2 A_{MIT}^R ~.
\end{equation}

\section{Computational model} \label{sec:model}
\subsection{Summary}
The wavefunctions of all Be-like ions ranging from boron ($Z=5$) to uranium ($Z=92$) are calculated using the latest version of the \textsc{Grasp2k} program suite \cite{Grasp2kV3} based on the MCDHF method briefly outlined above.

The radial parts of the Dirac-orbitals, together with the expansion coefficients, $c_i$, in Eq. (\ref{eq:csf_exp}) are optimized in a relativistic self-consistent field (RSCF) procedure based on the Dirac-Coulomb Hamiltonian. This part of the calculation is performed in a layer-by-layer scheme in which the active set of one-electron Dirac-orbitals is expanded systematically until satisfactory convergence of atomic properties, such as excitation energies, is achieved.

With a well-optimized basis at hand, the Breit interaction (in the low frequency limit) and leading QED effects are included in a subsequent relativistic configuration interaction (RCI) model. Both these effects grow in importance with increasing ionization stages as compared to electron correlation which becomes less important for high-$Z$ ions.

Finally, in order to calculate the transition rate according to Eq. (\ref{MIT}), the mixing coefficients $d_\mathcal{S}$ have to be evaluated. This is done using the first order perturbation theory approximation of Eq. (\ref{eq:mixingcoff}), with reduced matrix elements calculated using the \textsc{Grasp2k} module, \textsc{Hfszeeman} \cite{Andersson2008}.

\subsection{Optimization of Dirac-orbitals and electron correlation model}
The ASFs of the even, $2s^2~^1S_0$, and the odd parity states, $2s\,2p~^3P_{0,1,2},\,2s\,2p~^1P_1$, are determined in two separate calculations. The four odd ASFs are determined simultaneously in an extended optimal level (EOL) scheme \cite{Dyall1989CPC}, where the optimization is on a weighted sum of the corresponding fine structure energies. It should be noted that standard Racah algebra assumes the ASFs to be built from the same set of orthogonal radial orbitals. Thus to compute transition matrix elements between the even and odd parity ASFs, generated from independently optimized orbital sets, we apply biorthogonal transformation techniques to the orbital sets \cite{Olsen1995PRE,Jönsson1998PRA}, after which the calculation can be performed using standard methods. 

We use a correlation model in which the CSF space is generated using a complete active space (CAS) approach with orbitals up to $n=4$, and then merged with the result of single (S) and double (D) substitutions to higher $n$'s (with orbital angular momentum restricted by $l\le6$) from the multi-reference (MR) $\{2s^2,2p^2\}$ for the even parity states and the $\{2s\,2p\}$ reference for the odd states.

In order to capture as much correlation as possible in the computationally much less demanding RCI calculation, we extend the active space model from above by allowing also for triple (T) and quadruple (Q) substitutions with the restriction, in excess of the orbital angular momentum upper limit, that there should always be at-least two electrons in subshells with $n\le3$. This is in effect a simple way of generating a SD-expansion from a large MR.

The active space for ions with charge states $Z=5$ to $42$ is expanded up to $n=8$ according to the rules set up above, corresponding to a maximum (in the RCI calculation) of 37~653 and 296~215 CSFs of even ($J=0$) and odd parity ($J=0,1,2$) respectively. These calculations include 62 Dirac-orbitals. For $Z=43$ to $73$ it is sufficient with $n=7$, giving 23~205 even and 179~701 odd parity CSFs. For the highly charged ions we expect relativistic effects, Breit interaction and QED contributions to be far more important than correlation. For $Z=74$ to $85$ it is therefore sufficient with CSF expansions up to $n=6$, resulting in 12~541 even and 94~265 odd parity CSFs, and for $Z=86$ to $92$ we expand up to $n=5$ which corresponds to a maximum of 5~786 even and 41~723 odd parity CSFs.

\section{Results and Discussion} \label{sec:results}
\subsection{The magnetic field induced $2s\,2p~^3P_0~\to~2s^2~^1S_0$ E1 transition rates \label{sec:rates}}
{ \renewcommand{\arraystretch}{0.95} 
\begin{table}
\caption{\label{tab:MIT}Transition rates of the magnetic field induced E1 transition $2s^2~^1S_0 - 2s\,2p~^3P_0$. The reduced rates, $A_{MIT}^R$, of Eq. (\ref{eq:RMIT}), are given in two versions, the "Full" where both perturbers $^3P_1$ and $^1P_1$ are included, and "No $^1P_1$" where the $^1P_1$ perturber has been excluded. The  difference in percentage of these two approaches is presented in "$\delta_\%$". $A_{MIT}$ are rates for two example field strengths. Note that $[\#] = 10^\#$.}
\scriptsize{
\begin{tabular*}{\linewidth}{@{\extracolsep{\fill}}ccccccccccccccccccc}
\hline
\hline
& & & \multicolumn{3}{c}{$A_{MIT}^R(Z)\;[\mathrm{ s^{-1}T^{-2}}]$ (reduced)}  & & \multicolumn{2}{c}{$A_{MIT}(Z,B)\;[\mathrm{ s^{-1}}]$} & & & & & \multicolumn{3}{c}{$A_{MIT}^R\;[\mathrm{ s^{-1}T^{-2}}]$ (reduced)}  & & \multicolumn{2}{c}{$A_{MIT}(Z,B)\;[\mathrm{ s^{-1}}]$} \\  \cline{4-6} \cline{8-9}  \cline{14-16} \cline{18-19}
Ion  & $Z$ & & Full      & No $^1P_1$  &$\delta_\%$ & &  0.5 T  & 1.5 T &$\quad$ & Ion  & $Z$ & & Full & No $^1P_1$  &$\delta_\%$ & &  0.5 T & 1.5 T\\
\cline{1-9} \cline{11-19}
B    &5  &&4.078[-2]  &4.080[-2]  &0.04    &&1.020[-2] &9.176[-2]&&In   &49 &&3.371[-3]  &3.951[-3]  &17.21   &&8.426[-4]  &7.584[-3]\\
C    &6  &&2.697[-2]  &2.701[-2]  &0.12    &&6.743[-3] &6.069[-2]&&Sn   &50 &&3.361[-3]  &3.914[-3]  &16.46   &&8.402[-4]  &7.562[-3]\\
N    &7  &&2.081[-2]  &2.086[-2]  &0.24    &&5.203[-3] &4.683[-2]&&Sb   &51 &&3.356[-3]  &3.884[-3]  &15.73   &&8.390[-4]  &7.551[-3]\\
O    &8  &&1.746[-2]  &1.753[-2]  &0.43    &&4.364[-3] &3.928[-2]&&Te   &52 &&3.353[-3]  &3.856[-3]  &15.02   &&8.382[-4]  &7.544[-3]\\
F    &9  &&1.500[-2]  &1.510[-2]  &0.69    &&3.749[-3] &3.374[-2]&&I    &53 &&3.355[-3]  &3.836[-3]  &14.34   &&8.388[-4]  &7.549[-3]\\
Ne   &10 &&1.318[-2]  &1.331[-2]  &1.03    &&3.294[-3] &2.965[-2]&&Xe   &54 &&3.358[-3]  &3.817[-3]  &13.68   &&8.395[-4]  &7.555[-3]\\
Na   &11 &&1.177[-2]  &1.195[-2]  &1.48    &&2.943[-3] &2.649[-2]&&Cs   &55 &&3.366[-3]  &3.805[-3]  &13.05   &&8.415[-4]  &7.573[-3]\\
Mg   &12 &&1.065[-2]  &1.087[-2]  &2.02    &&2.663[-3] &2.397[-2]&&Ba   &56 &&3.375[-3]  &3.795[-3]  &12.44   &&8.438[-4]  &7.595[-3]\\
Al   &13 &&9.739[-3]  &1.000[-2]  &2.68    &&2.435[-3] &2.191[-2]&&La   &57 &&3.389[-3]  &3.791[-3]  &11.86   &&8.472[-4]  &7.625[-3]\\
Si   &14 &&8.980[-3]  &9.291[-3]  &3.46    &&2.245[-3] &2.021[-2]&&Ce   &58 &&3.405[-3]  &3.791[-3]  &11.31   &&8.513[-4]  &7.662[-3]\\
P    &15 &&8.340[-3]  &8.703[-3]  &4.36    &&2.085[-3] &1.876[-2]&&Pr   &59 &&3.425[-3]  &3.795[-3]  &10.78   &&8.563[-4]  &7.707[-3]\\
S    &16 &&7.793[-3]  &8.212[-3]  &5.38    &&1.948[-3] &1.754[-2]&&Nd   &60 &&3.448[-3]  &3.802[-3]  &10.28   &&8.620[-4]  &7.758[-3]\\
Cl   &17 &&7.321[-3]  &7.798[-3]  &6.51    &&1.830[-3] &1.647[-2]&&Pm   &61 &&3.473[-3]  &3.813[-3]  &9.80    &&8.682[-4]  &7.814[-3]\\
Ar   &18 &&6.909[-3]  &7.445[-3]  &7.77    &&1.727[-3] &1.555[-2]&&Sm   &62 &&3.501[-3]  &3.828[-3]  &9.35    &&8.751[-4]  &7.876[-3]\\
K    &19 &&6.547[-3]  &7.144[-3]  &9.12    &&1.637[-3] &1.473[-2]&&Eu   &63 &&3.532[-3]  &3.846[-3]  &8.91    &&8.829[-4]  &7.946[-3]\\
Ca   &20 &&6.227[-3]  &6.885[-3]  &10.56   &&1.557[-3] &1.401[-2]&&Gd   &64 &&3.564[-3]  &3.867[-3]  &8.50    &&8.911[-4]  &8.020[-3]\\
Sc   &21 &&5.943[-3]  &6.661[-3]  &12.07   &&1.486[-3] &1.337[-2]&&Tb   &65 &&3.602[-3]  &3.894[-3]  &8.10    &&9.005[-4]  &8.105[-3]\\
Ti   &22 &&5.689[-3]  &6.464[-3]  &13.63   &&1.422[-3] &1.280[-2]&&Dy   &66 &&3.643[-3]  &3.925[-3]  &7.73    &&9.108[-4]  &8.197[-3]\\
V    &23 &&5.460[-3]  &6.290[-3]  &15.21   &&1.365[-3] &1.228[-2]&&Ho   &67 &&3.684[-3]  &3.955[-3]  &7.37    &&9.209[-4]  &8.288[-3]\\
Cr   &24 &&5.253[-3]  &6.135[-3]  &16.78   &&1.313[-3] &1.182[-2]&&Er   &68 &&3.734[-3]  &3.997[-3]  &7.03    &&9.335[-4]  &8.402[-3]\\
Mn   &25 &&5.067[-3]  &5.994[-3]  &18.31   &&1.267[-3] &1.140[-2]&&Tm   &69 &&3.785[-3]  &4.039[-3]  &6.71    &&9.463[-4]  &8.516[-3]\\
Fe   &26 &&4.893[-3]  &5.861[-3]  &19.78   &&1.223[-3] &1.101[-2]&&Yb   &70 &&3.838[-3]  &4.083[-3]  &6.40    &&9.594[-4]  &8.635[-3]\\
Co   &27 &&4.743[-3]  &5.745[-3]  &21.13   &&1.186[-3] &1.067[-2]&&Lu   &71 &&3.894[-3]  &4.132[-3]  &6.11    &&9.735[-4]  &8.761[-3]\\
Ni   &28 &&4.602[-3]  &5.631[-3]  &22.36   &&1.151[-3] &1.035[-2]&&Hf   &72 &&3.951[-3]  &4.181[-3]  &5.83    &&9.877[-4]  &8.889[-3]\\
Cu   &29 &&4.474[-3]  &5.523[-3]  &23.44   &&1.118[-3] &1.007[-2]&&Ta   &73 &&4.010[-3]  &4.233[-3]  &5.57    &&1.002[-3]  &9.022[-3]\\
Zn   &30 &&4.357[-3]  &5.418[-3]  &24.35   &&1.089[-3] &9.803[-3]&&W    &74 &&4.315[-3]  &4.544[-3]  &5.31    &&1.079[-3]  &9.708[-3]\\
Ga   &31 &&4.249[-3]  &5.315[-3]  &25.08   &&1.062[-3] &9.561[-3]&&Re   &75 &&4.395[-3]  &4.618[-3]  &5.07    &&1.099[-3]  &9.889[-3]\\
Ge   &32 &&4.151[-3]  &5.214[-3]  &25.62   &&1.038[-3] &9.339[-3]&&Os   &76 &&4.477[-3]  &4.694[-3]  &4.84    &&1.119[-3]  &1.007[-2]\\
As   &33 &&4.060[-3]  &5.115[-3]  &25.97   &&1.015[-3] &9.136[-3]&&Ir   &77 &&4.565[-3]  &4.776[-3]  &4.62    &&1.141[-3]  &1.027[-2]\\
Se   &34 &&3.977[-3]  &5.017[-3]  &26.13   &&9.944[-4] &8.949[-3]&&Pt   &78 &&4.656[-3]  &4.861[-3]  &4.41    &&1.164[-3]  &1.048[-2]\\
Br   &35 &&3.902[-3]  &4.921[-3]  &26.13   &&9.755[-4] &8.779[-3]&&Au   &79 &&4.751[-3]  &4.951[-3]  &4.21    &&1.188[-3]  &1.069[-2]\\
Kr   &36 &&3.832[-3]  &4.827[-3]  &25.97   &&9.581[-4] &8.623[-3]&&Hg   &80 &&4.846[-3]  &5.041[-3]  &4.03    &&1.211[-3]  &1.090[-2]\\
Rb   &37 &&3.769[-3]  &4.736[-3]  &25.67   &&9.422[-4] &8.480[-3]&&Tl   &81 &&4.945[-3]  &5.135[-3]  &3.84    &&1.236[-3]  &1.113[-2]\\
Sr   &38 &&3.712[-3]  &4.649[-3]  &25.25   &&9.279[-4] &8.351[-3]&&Pb   &82 &&5.046[-3]  &5.231[-3]  &3.67    &&1.262[-3]  &1.135[-2]\\
Y    &39 &&3.660[-3]  &4.564[-3]  &24.72   &&9.149[-4] &8.234[-3]&&Bi   &83 &&5.151[-3]  &5.332[-3]  &3.51    &&1.288[-3]  &1.159[-2]\\
Zr   &40 &&3.614[-3]  &4.485[-3]  &24.11   &&9.034[-4] &8.131[-3]&&Po   &84 &&5.261[-3]  &5.437[-3]  &3.35    &&1.315[-3]  &1.184[-2]\\
Nb   &41 &&3.572[-3]  &4.409[-3]  &23.43   &&8.930[-4] &8.037[-3]&&At   &85 &&5.370[-3]  &5.542[-3]  &3.20    &&1.343[-3]  &1.208[-2]\\
Mo   &42 &&3.533[-3]  &4.335[-3]  &22.71   &&8.833[-4] &7.949[-3]&&Rn   &86 &&5.451[-3]  &5.617[-3]  &3.06    &&1.363[-3]  &1.226[-2]\\
Tc   &43 &&3.495[-3]  &4.263[-3]  &21.96   &&8.738[-4] &7.864[-3]&&Fr   &87 &&5.555[-3]  &5.718[-3]  &2.92    &&1.389[-3]  &1.250[-2]\\
Ru   &44 &&3.466[-3]  &4.200[-3]  &21.17   &&8.665[-4] &7.799[-3]&&Ra   &88 &&5.641[-3]  &5.799[-3]  &2.79    &&1.410[-3]  &1.269[-2]\\
Rh   &45 &&3.441[-3]  &4.142[-3]  &20.37   &&8.602[-4] &7.742[-3]&&Ac   &89 &&5.737[-3]  &5.890[-3]  &2.66    &&1.434[-3]  &1.291[-2]\\
Pd   &46 &&3.419[-3]  &4.088[-3]  &19.57   &&8.548[-4] &7.693[-3]&&Th   &90 &&5.816[-3]  &5.964[-3]  &2.54    &&1.454[-3]  &1.309[-2]\\
Ag   &47 &&3.401[-3]  &4.040[-3]  &18.77   &&8.503[-4] &7.653[-3]&&Pa   &91 &&5.914[-3]  &6.057[-3]  &2.43    &&1.478[-3]  &1.331[-2]\\
Cd   &48 &&3.385[-3]  &3.994[-3]  &17.99   &&8.464[-4] &7.617[-3]&&U    &92 &&5.978[-3]  &6.116[-3]  &2.32    &&1.494[-3]  &1.345[-2]\\
\hline
\hline
\end{tabular*} }
\end{table}
}

Magnetic field induced rates of the $2s^2~^1S_0 - 2s\,2p~^3P_0$ transition for Be-like ions, in a comparatively weak magnetic field can be estimated from the reduced transition rates, $A_{MIT}^R$, as defined in Eq. (\ref{eq:RMIT}). Using this method, we calculate rates for all ions in the beryllium isoelectronic sequence with zero nuclear spin, between $Z=5$ and $Z=92$. 
In these calculations the wave function of the $2s\,2p~^3P_0$ state under the influence of an external magnetic field, is approximately described including $2s\,2p~^3P_1$ and $2s\,2p~^1P_1$ as perturbers. The resulting reduced rates are presented in Table \ref{tab:MIT} in the column labelled "Full" (since both perturbers are included). It is found that the MIT rates are small and almost constant ($\sim 3\times10^{-3}\mathrm{s^{-1}}$) for high-$Z$ ions. These rates will be compared to the expected E1M1 rates in the subsequent section. Note that the reduced MIT rates by definition correspond to an external magnetic field strength of $1$ T.

We also investigate the importance of the $2s\,2p~^1P_1$ perturber by comparing our results to a calculation, labelled "No $^1P_1$" Table \ref{tab:MIT}, where we only include $2s\,2p~^3P_1$. The next collumn in this table, labelled "$\delta_\%$", shows the difference between these two approaches in percentage. It is clear that excluding $^1P_1$ leads to a significant overestimation of the rates, by more than $5\%$ for $Z=16$ to $75$, reaching as much as $26 \%$ for Be-like Se and Br. The full calculation reduces the MIT rates, as compared to only including $^1P_1$, due to cancellation of the individual transition amplitudes from the two perturbers involved in Eq. (\ref{MIT}). 

The two remaining columns of Table \ref{tab:MIT} (labelled "$B$") show two example field strengths of $0.5$ and $1.5$ T respectively, where the latter is the maximum field strength of the dipole bending magnets of ESR as mentioned in section \ref{sec:rings}. Note that the bending magnets only cover parts of the storge-ring, resulting in a smaller effective field. Details on results connected to the particular experiment suggested at ESR is presented in subsection \ref{sec:ESR}.

\subsection{Uncertainties of results \label{sec:check}} 
In order to benchmark the quality of our wavefunctions (and ultimately the magnetic induced transition rates) we compare our obtained excitation energies, energy separations between the reference state $^3P_0$ and the perturbers $^{1,3}P_{1}$, as-well as the involved $J$-dependent transition rates with experiment \cite{RyabtsevPSc2005} and another accurate theoretical work based on B-spline relativistic configuration interaction calculations \cite{ChengPRA2008} with a careful treatment of QED effects (from here on this calculation will be referred to as BSRCI for brevity). In this subsection we also present results from convergence studies of the involved parameters, the influence of QED effects and finally we investigate the impact of neglecting additional perturbers.

This initial comparison with experiment and other theory is presented in Table \ref{tab:comparision} for three selected charge states representing the neutral end, the intermediate and the highly charged ions, which allows us to test our model with respect to correlation, relativistic and QED effects. Note that the excitation energy of $^3P_1$ and $^1P_1$ are, in difference from the other parameters presented, of no direct importance in the evaluation of the MIT rates. On the other hand, the accuracy of e.g. the energy separations between the reference state and the perturbers are of particular interest as they enter directly in the evaluation of the magnetic mixing coefficients $d_{\Gamma J}$ in Eq. (\ref{eq:mixingcoff}). The same goes for the E1 transition matrix elements of $^3P_1 \to {^1S_0} $ and $^1P_1 \to {^1S_0}$.

The first ion, chosen to represent the neutral end in this test, is singly ionized boron ($Z=5$). We expect energies of ions this close to neutral to be dominated by contributions from electron correlation (relative to Dirac-Fock energies). It can be seen from Table \ref{tab:comparision} that both excitation and separation energies are in excellent agreement with experiment, in most cases well within $1\%$. Tin ($Z=50$) is chosen to represent the intermediate part of the sequence. This should be a simple calculation in our approach as both correlation and QED effects are small and other relativistic effects taken care of efficiently, which is apparent from the resulting energies as the difference from BSRCI is almost negligible. For the high end of the sequence we choose the ion of highest charge state in this work, namely uranium ($Z=92$). There is a slightly larger disagreement between our excitation energies and the BSRCI results for the intermediate ions (as expected due to a greater influence from QED effects) with a maximum deviation of $1.22\%$. The energy separations however, are in very good agreement with the BSRCI energies. 

We continue with the convergence of the MIT rates, and the involved parameters, as the active set of orbitals is increased according the model presented in section \ref{sec:model}. We conclude from our studies that the MIT rates are converged to $5\%$ on the far neutral end of the sequence, whereas the rates for the highly charged ions are almost fully converged, as expected.

The QED contribution to the MIT rate ranges from zero in the neutral end, to $-15\%$ for $Z=50$ and about $-40\%$ for $Z=92$. In our relatively simple QED model, we estimate an upper limit of the QED uncertainty for the mid-$Z$ ions, to about $5\%$. As the impact of QED is almost half of the total rate for the highly charged ions, the errors are most certainly larger as well. This is however of no direct problem to this work as the E1M1 rate is anyway dominating over the MIT rate for these ions by more than two orders of magnitude for Z=75 and up (see Fig. \ref{fig:E1M1_MIT} in the following subsection). Note however that the formula for the E1M1 rate by Laughlin, Eq. \eqref{eq:E1M1} \cite{LaughlinPLA1980}, which is used in this work for reference values, is non-relativistic and ignores the $^1P_1$ state and should therefore be used with care, especially for the highly charged ions. The competition between the MIT and E1M1 rates will be discussed further in the following subsection (\ref{sec:comp}).

It remains to estimate the influence of including perturbers other than $2s\,2p~^1P_1$ and $2s\,2p~^3P_1$. Starting with $Z=5$, we evaluate the size of the magnetic field induced mixing, Eq. \eqref{eq:mixingcoff}, with the closest lying energy level of odd parity above the $2s\,2p~^1P_1$ level. This is the state $2s\,3p~^3P_1$ having an energy separation with the $2s\,2p~^3P_0$ reference state of $106~655~\mathrm{cm^{-1}}$ according to the NIST ASD \cite{NIST}. Furthermore, the matrix element involved in the evaluation of the mixing coefficient is about a factor of $10$ smaller for $2s\,3p~^3P_1$ which implies a mixing coefficient in total 30 times smaller than the one with $2s\,2p~^1P_1$. The same line of action for $Z=50$ gives a mixing coefficient with $2s\,3p~^3P_1$ that is more than 1500 times smaller than the one for $2s\,2p~^1P_1$. Calculating the MIT rates including the $2s\,3p$ perturbers reveals an additional contribution of about $0.01\%$ for $Z=92$, to zero for $Z=5$, and we conclude that neglecting further perturbers has a negligible impact on the MIT rates.

We end the discussion of this subsection by giving an estimation of the total MIT transition rate uncertainty. After a careful convergence study we concluded the rates to be converged to within $5\%-0\%$ from low- to high-$Z$ ions. We estimate the accuracy of the QED contribution for the low- to mid-Z ions to $0\%-5\%$ and around $10\%$ for the highly charged ions. The influence of neglecting further perturbers is very small, and thus we estimate the total accuracy of our MIT rates to be varying from about $5\%$ for low- and mid-Z ions, up to $10\%$ as an upper limit for the highly charged ions, where should be clear that the errors related to QED are very hard to evaluate.
\begin{table}[t] 
\caption{\label{tab:comparision}Comparison of excitation energies, $E$, energy separations with the $^3P_0$ reference state, $\Delta E$,  and transition rates, $A$, involved in the MIT calculation for some selected charge states. We compare with experimental energies \cite{RyabtsevPSc2005} and theoretical rates \citep{Tachiev1999} from the NIST database \cite{NIST} for $Z=5$. Results for ions of charge states $Z=50$ and $92$ is compared with another recent accurate theoretical work \cite{ChengPRA2008} (labelled "BSRCI"). The fractional difference relative to our results are given in rows labelled by "$\delta_{\%}$". All energies are given in units of $\mathrm{cm^{-1}}$, transition rates in $s^{-1}$ and numbers in brackets represent powers of ten. Note that our values are rounded off to the same number of digits as given in the data compared with.}
\begin{footnotesize}
\begin{tabular*}{1.00\linewidth}{@{\extracolsep{\fill}}rrcrrrrrrrr}
\hline \hline
ion & source & & \multicolumn{1}{c}{$E_{\,^3P_0}$ }& \multicolumn{1}{c}{$\Delta E_{\,^3P_0-^3P_1}$}  & \multicolumn{1}{c}{$E_{\,^3P_1}$ }  &\multicolumn{1}{c}{$\Delta E_{\,^3P_0-^1P_1}$} & \multicolumn{1}{c}{$E_{\,^1P_1}$}  & & \multicolumn{1}{c}{$A_{\,^3P_1\to^1S_0}$} & \multicolumn{1}{c}{$A_{\,^1P_1\to^1S_0}$} \\ 
\cline{1-2} \cline{4-8} \cline{10-11}

\multirow{3}{*}{B$_{Z=5}  $  \Bigg \{ }
                                     & this work
                                     & & 3.727186[4] & 6.06[0] & 3.727792[4] &  3.62225[4] & 7.349438[4]  & & 1.03[1] & 1.20[9] \\
                                     & NIST
                                     & & 3.733554[4] & 6.11[0] & 3.734165[4] &  3.60610[4]  & 7.339651[4] & & 1.04[1] & 1.20[9]\\
                                     & $\delta_{\%}$
                                     & & $0.17$      & $0.82$  & $0.17$      &  $-0.45$     & $-0.13$     & & $0.97$  & $0.00$ \\
\multirow{3}{*}{Sn$_{Z=50}$ \Bigg \{ }
                                     & this work
                                     & & 7.58328[5]  & 1.63389[5] & 9.21716[5] &  2.5829[6]  & 3.34121[6] & & 1.652[9] & 3.819[11] \\
                                     & BSRCI
                                     & & 7.58449[5]  & 1.63306[5] & 9.21755[5] &  2.5807[6] & 3.33919[6]  & & 1.676[9] & 3.825[11] \\
                                     & $\delta_{\%}$
                                     & & $0.02$      & $-0.05$    & $0.00$     &  $-0.09$   & $-0.06$     & & $1.45$   & $0.16$    \\
\multirow{3}{*}{U$_{Z=92}$ \Bigg \{ }& this work
                                     & & 2.10309[6]  & 3.2251[5]  & 2.42561[6] &  3.4277[7] & 3.63804[7]  & & 9.806[9] & 1.169[14]\\
                                     & BSRCI
                                     & & 2.07739[6]  & 3.2241[5]  & 2.39980[6] &  3.4230[7] & 3.63083[7]  & & 9.773[9] & 1.166[14] \\
                                     & $\delta_{\%}$
                                     & &$-1.22$      & $-0.03$    & $-1.06$    &  $-0.14$   & $-0.20$     & & $−0.34$  & $-0.26$\\
\hline \hline
\end{tabular*}
\end{footnotesize}
\end{table}

\subsection{The competition between the E1M1 two-photon and MIT decays \label{sec:comp}}
\begin{figure}[b]
\includegraphics[height=0.25\textheight]{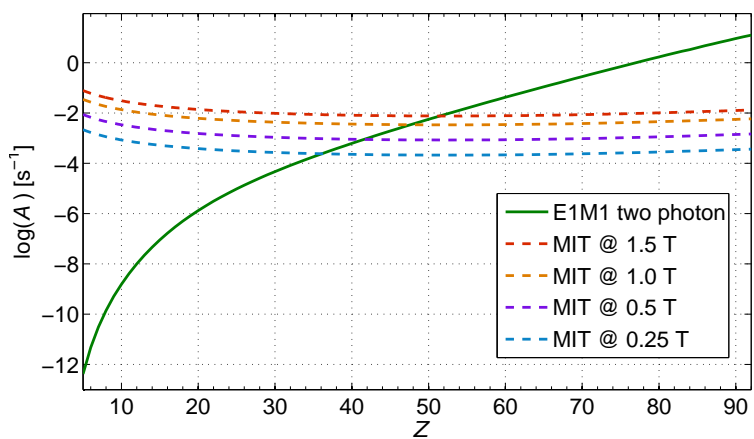}
\vspace{-15pt}
\caption{\label{fig:E1M1_MIT} Comparison of transition rates for the two different decay channels of the $2s\,2p~^3P_0~\to~2s^2~^1S_0$ transition in Be-like atoms with zero nuclear spin under influence of an external magnetic field. The dashed lines shows how the magnetic induced transition (MIT) rate varies along the isoelectronic sequence, and the green solid line shows the behaviour of the E1M1 two-photon transition rate as predicted by Eq. (\ref{eq:E1M1}). The rates are given in logarithmic scale.}
\end{figure}
To be able to extract the E1M1 two-photon transition rate, one needs to determine, or at least estimate the MIT rate, which was done in subsection \ref{sec:rates}. In order to predict the influence of the magnetic field on the total lifetime, one should compare these MIT rates to the corresponding two-photon transition rates. We begin, however, by evaluating the MIT rates further, along the sequence.

\begin{figure}[b]
\includegraphics[height=0.25\textheight]{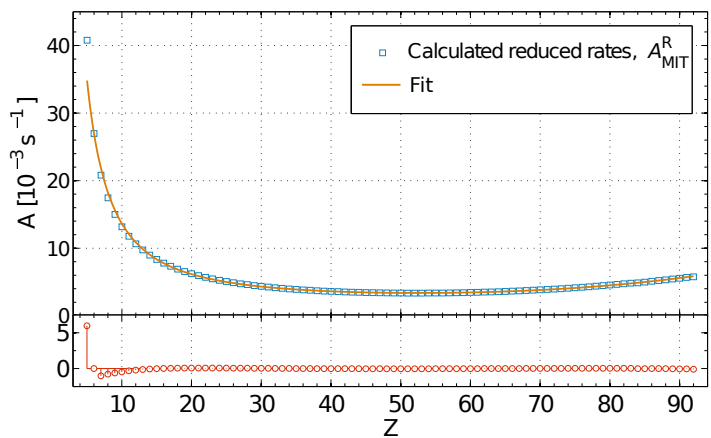}
\vspace{-15pt}
\caption{\label{fig:fit}Reduced transition rates of the magnetic induced transition $2s\,2p~^3P_0 \to 2s^2~^1S_0$ due to mixing with the intercombination channel, see Fig. \ref{fig:1}, which in turn is allowed due to mixing with the $^1P_1$ level. The solid line shows the least squares fit presented in Eq. (\ref{eq:fit}). The residuals of the fit are shown in the lower panel of the figure.}
\end{figure}

According to the scaling law for physical quantities in the hydrogenic approximation, the MIT rate is roughly proportional to $Z^4$ for high-Z ions. To obtain better description of the dependence of the MIT rate on the atomic number, we perform a non-linear least squares fit of the calculated reduced rates including an extra general term of $Z$, resulting in the following expression
\begin{equation} \label{eq:fit}
A_{MIT}^R(Z) = \alpha Z^\delta + \beta Z^4 + \gamma
\end{equation}
where 
\begin{eqnarray} \label{eq:fit_param}
\alpha =  3.703\times 10^{-1}\mathrm{ s^{-1}T^{-2}} & ~ , \quad & \beta  =  4.717\times 10^{-11}\mathrm{ s^{-1}T^{-2}} \nonumber \\
\gamma =  2.074\times 10^{-3}\mathrm{ s^{-1}T^{-2}} & ~ , \quad & \delta = -1.507\qquad \qquad \qquad \quad .
\end{eqnarray}
The first term in Eq. (\ref{eq:fit}) dominates at the neutral and intermediate part of the sequence and accounts for the deviation from the hydrogenic behaviour and corresponds to e.g. stronger correlation effects. The $Z^4$-dependence increasingly makes up the major part of the rate from about $Z=70$ and up as expected from the hydrogenic scaling laws. A plot of the computed MIT rate data points of Table \ref{tab:MIT}, and the fitted curve is presented in Fig. \ref{fig:fit} together with the deviation of the fit in the panel underlying the main plot. As can be seen from this figure, the fit is in excellent agreement with the calculations, except for a few ions at the neutral end of the sequence where the reduced transition rates of Table \ref{tab:MIT} should be used instead if there is a need for high accuracy.

Using Eq. (\ref{eq:RMIT}) we can write down a final formula for the non-reduced MIT rate, as
\begin{equation} \label{eq:fit_tot}
A_{MIT}(Z,B) = A_{MIT}^R(Z)B^2 = \left[\alpha Z^\delta+\beta Z^4+\gamma\right]B^2
\end{equation}
with the constants $\alpha,~\beta,~\gamma$ and $\delta$ given by Eq. (\ref{eq:fit_param}) above.

For the E1M1 decay from $2s\,2p~^3P_0$ in Be-like ions only two theoretical predictions by Schmieder \cite{SchmiederPRA1973} and by Laughlin \cite{LaughlinPLA1980} are available. As mentioned in the previous subsection, Laughlin's derivation is non-relativistic and ignores the $2s\;2p~^1P_1$ state. Thus should all comparisons between the MIT and E1M1 transition rates in this subsection, be read with the knowledge that the accuracy of the E1M1 rates could be very low for high-Z ions. Bernhardt \textit{et al.} \cite{Bernhardt2012} evaluated the integrals involved in Laughlins's expression for the rate analytically, with the resulting formula
\begin{eqnarray} \label{eq:E1M1}
A_{E1M1}(E,\Delta,Z) & = \frac{1}{6} A_0 Z^4 \Bigg[& E^5 - 8E^4\Delta - 68E^3\Delta^2 - 120E^2\Delta^3 - 60E\Delta^4 \nonumber \\
                   &   & ~ + \frac{12\Delta^2 (3E^2+10E\Delta + 10\Delta^2)(E+\Delta)^2}{E+2\Delta}\ln\Big(\frac{E+\Delta}{\Delta}\Big) \Bigg]
\end{eqnarray}
where $E$ here represents the excitation energy of the $2s\,2p~^3P_0$ state, $\Delta$ is the $2s\,2p~^3P_{0,1}$ fine structure splitting and $A_0=2.867\times10^{-11}\mathrm{s^{-1}}$.

In Fig. \ref{fig:E1M1_MIT} we illustrate the E1M1 two-photon transition rates along the Be-like sequence according to this formula. In the same figure we also include the MIT rates calculated with some examples of magnetic field strengths between $B=0.25$ and $1.5$ T. From this figure it can be seen that the rates of the two transition channels are in general of comparable size for mid-$Z$ ions. At the low-$Z$ end of the sequence the MIT becomes the dominating decay channel, while for high-$Z$ ions the E1M1 channel has a much faster rate. 
\begin{figure}[t]
\includegraphics[height=0.25\textheight]{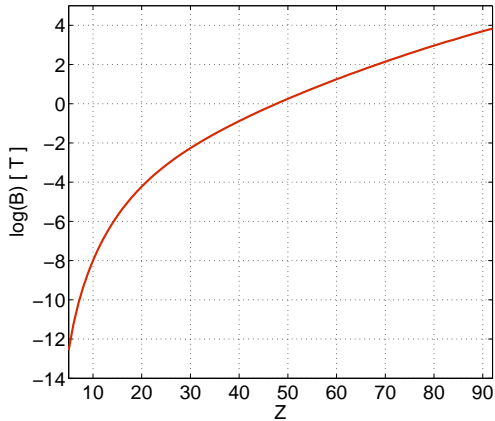}
\vspace{-15pt}
\caption{\label{fig:critical} A plot of the critical magnetic field strength, $B^{critical}$, as defined in Eq. \eqref{eq:critical}, along the Be-like sequence in log-scale. These field strengths correspond an MIT rate, $A_{MIT}$, equal to the E1M1 two-photon rate, $A_{E1M1}$.}
\end{figure}
Furthermore, we define a critical magnetic field strength, $B^{critical}$, as the field corresponding to a MIT rate equal to the E1M1 rate, leading to the following relation
\begin{equation} \label{eq:critical}
B^{critical} \equiv \sqrt{\frac{A_{E1M1}}{A_{MIT}^R}} \quad .
\end{equation}
Using Eq. \eqref{eq:E1M1} for the E1M1 rates (well-aware of the possible low accuracy of the E1M1 transition rates for the highly charged ions as pointed out earlier) and our reduced MIT rates from Table \ref{tab:MIT}, we calculate the critical magnetic field strength along the sequence, see Fig. \ref{fig:critical}. From this figure one can readily estimate in which region of $Z$ and $B$ the impact of the magnetic field on the lifetime of $^3P_0$ is of significant importance. For example, for Sn, $Z=50$, the critical magnetic field is about $0.3$ T. Hence it is crucial to include and evaluate the MIT rate in measurements involving an external magnetic field of this magnitude.

\subsection{The E1M1 transition rate measurement at ESR \label{sec:ESR}}
Considering the particular case of the proposed experiment at ESR, the six dipole bending magnets cover $36.24\%$ of the ring as explained in section \ref{sec:rings}. Each one of these magnets have a constant and near homogeneous magnetic field that can be set to any field strength, $B^{bending}_{magnets}$, up to a maximum of $1.5$ T, depending on the mass and energy of the stored ion. Hence we may conclude that, using Eq. (\ref{eq:fit_tot}), the effective rate of the $2s\,2p~^3P_0~\to~2s^2~^1S_0$ transition in a measurement at ESR can be estimated from
\begin{eqnarray}
A_{eff}^{ESR}(Z,B_{eff}) & = & A_{MIT}^{ESR}(Z,B_{eff}) + A_{E1M1}(Z) \\ 
                   & = & {B_{eff}}^2 A_{MIT}^R(Z) + A_{E1M1}(Z) \\ 
                   & = & {B_{eff}}^2\left[\alpha Z^\delta+\beta Z^4+\gamma\right] + A_{E1M1}(Z) \label{eq:conc2}
\end{eqnarray}
where $B_{eff}=\sqrt{0.3624} \times B^{bending}_{magnets}$ is the effective magnetic field strength deduced from averaging the MIT rate over one full revolution of ESR. The effective field strength is variable up to a maximum of $0.90$ T.

One can then choose to use either the reduced transition rates, $A_{MIT}^R$, given in Tab. \ref{tab:MIT}, or just simply the $Z$-dependent function in square brackets of Eq. (\ref{eq:conc2}) with fitted parameters defined in Eq. (\ref{eq:fit_param}). The measured total lifetime of $^3P_0$ can then be directly related to this rate, from which an approximation of the E1M1 transition rate can be obtained.

\section{Conclusions \label{sec:conc}}
In summary, it has been proposed \cite{Schippers2011c,Bernhardt2012,SchippersArXiv2012} to measure the rate of the $2s\,2p~^3P_0~\to~2s^2~^1S_0$ E1M1 two-photon transition in heavy Be-like ions with zero nuclear spin at the heavy-ion storage-ring ESR of the GSI Helmholtz Center for Heavy-Ion Research in Darmstadt, Germany. The E1M1 two-photon transition is the lowest order transition for Be-like isotopes with zero nuclear spin in a field-free environment. In a storage-ring however, the bending magnets generate magnetic fields which possibly could have a large impact on the lifetime of the $^3P_0$ level through magnetic field quenching. In this work we therefore present accurate theoretical transition rates of the magnetic field induced transition $2s\,2p~^3P_0~\to~2s^2~^1S_0$ in Be-like ions with the purpose of aiding such storage-ring measurements. 

Our theoretical approach is based on accurate wavefunctions calculated in an MCDHF scheme followed by a large-scale RCI calculation, as described in sections \ref{sec:method} and \ref{sec:model}. The MIT rates can then be obtained through Eq. (\ref{MIT}) and we present results and discussions in section \ref{sec:results}. The transition rates are presented in Table \ref{tab:MIT}, but also as a function of the charge state, $Z$, of the ion of interest and the magnetic field strength, $B$ (see Eq. (\ref{eq:fit_tot})). The quality of our results is motivated through a comparison with experiments and other theoretical results as well as convergence studies, the impact of QED effects and an evaluation of the influence of neglecting further perturbers. We conclude our errors to be within $5\%$ for low- and mid-$Z$ and slightly higher for the highly charged due to larger QED effects.

We continue by investigating how big impact the external $B$-field would have on the total measured lifetime of the $^3P_0$ level as compared to the lifetime associated with the E1M1 two-photon transition. In Fig. \ref{fig:E1M1_MIT} an approximate theoretical prediction of the E1M1 transition rate, Eq. (\ref{eq:E1M1}), is compared to our calculated MIT rates for some different typical field strengths. The figure shows that storage-ring measurements in general, involving external magnetic fields, is infeasible for ions at the neutral end of the isoelectronic sequence where the MIT channel is completely dominating. But foremost we can conclude that in order to determine the E1M1 rate from such a measurements, an accurate evaluation of the MIT rates is crucial for ions around $Z=25$ to $65$, where the MIT and E1M1 transition rates are of the same order of magnitude.

Finally we apply our results to the particular measurement proposed at ESR. We introduce an effective magnetic field, $B_{eff}$, depending on the bending magnet set-up of ESR, and give a relation, Eq. \eqref{eq:conc2}, for the total transition rate corresponding to the measured lifetime of $^3P_0$. From this expression one can then, if the total transition rate can be evaluated experimentally, readily obtain an estimation of the rate associated with the E1M1 two-photon transition.

\begin{acknowledgments}
J.G., J.L., T.B. and P.J. gratefully acknowledge support from the Swedish Research Council (Vetenskapsr{\aa}det) and the Swedish Institute under the Visby-program. D.B. and S.S. were supported by Deutsche Forschungsgeinschaft (DFG) under contract no. Schi378/8-1.
\end{acknowledgments}

\bibliography{refs}

\end{document}